\documentclass[aps,prd,longbibliography,superscriptaddress,twocolumn,showpacs,showkeys,floatfix,a4paper,eqsecnum]{revtex4-1}

\usepackage{datetime}
\usepackage{changes}
\usepackage{textcomp}
\usepackage{eurosym}
\usepackage{amsfonts}
\usepackage{array}
\usepackage{amsthm}
\usepackage{bm}
\usepackage{palatino}
\usepackage[colorinlistoftodos]{todonotes}
\usepackage{mathpazo}
\usepackage{supertabular}
\usepackage{subfig}
\usepackage{amssymb}
\usepackage{eurosym}
\usepackage{amsmath}
\usepackage{epsfig}
\usepackage{graphics}
\usepackage{color}
\usepackage{graphicx}
\usepackage[font={footnotesize,it}]{caption}
\usepackage[colorlinks=true,
            linkcolor=blue,
            urlcolor=blue,
            citecolor=blue]{hyperref}
\def\be{\begin{equation}}
\def\ee{\end{equation}}
\def\beq{\begin{eqnarray}}
\def\eeq{\end{eqnarray}}

\def\bes{\begin{eqnarray}}
\def\ees{\end{eqnarray}}

\def\LOGO{%
\begin{picture}(0,0)\unitlength=1cm
\put (3,-1) {\includegraphics[width=5em]{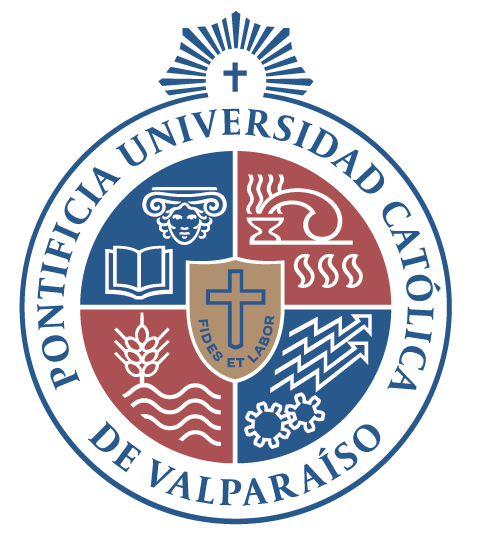}}
\end{picture}
}

\begin{document}

\begin{center}
  \sffamily\bfseries
  {\Large }\LOGO
\end{center}

\title{Light deflection by Damour-Solodukhin wormholes and Gauss-Bonnet theorem}

\author{Ali \"{O}vg\"{u}n}
\email{ali.ovgun@pucv.cl}
\homepage{http://www.aovgun.com}
\affiliation{Instituto de F\'{\i}sica, Pontificia Universidad Cat\'olica de
Valpara\'{\i}so, Casilla 4950, Valpara\'{\i}so, Chile.}

\affiliation{Physics Department, Arts and Sciences Faculty, Eastern Mediterranean University, Famagusta, North Cyprus via Mersin 10, Turkey.}

\affiliation{School of Natural Sciences, Institute for Advanced Study, 1 Einstein Drive
Princeton, NJ 08540, USA.}

\date{\today }

\begin{abstract}
In this paper, using the recent method proposed by Ono,
Ishihara and Asada (OIA) who extend the idea of Gibbons and Werner to
the stationary and axisymmetric case, we apply the Gauss-Bonnet theorem to the optical metric of the non-rotating and rotating Damour-Solodukhin wormholes spacetimes to study the weak gravitational lensing by these objects. Furthermore, we study the strong gravitational lensing by the non-rotating Damour-Solodukhin wormholes using the Bozza's method to see the differences between the weak lensing and the strong lensing. We demonstrate the relation between the strong deflection angle and quasinormal modes of the Damour-Solodukhin wormholes. Interestingly it is found that the wormhole parameter $\lambda$, affects the deflection of light in strong and weak limits compared to the previous studies of gravitational lensing by Schwarzschild black holes. Hence, the results provide a unique tool to shed light on the possible existence of wormholes. 
\end{abstract}

\maketitle

\section{Introduction}
Einstein-Rosen (ER) bridge is a consequence of Einstein's theory of relativity similarly to black holes. The ER equation glues to distant points of spacetime. This was firstly introduced by Einstein and Rosen in 1935 and then ER is refereed to as a wormhole \cite{Einstein:1935tc}. On the other hand, Morris and Thorne in 1988 showed that constructing traversable wormhole solution is also possible, however it costs to necessity of exotic matter \cite{Morris:1988cz,Morris:1988tu}. Afterwards, inspiring by the Morris-Thorne paper, many physicists study wormhole in different aspects \cite{Damour:2007ap,Bueno:2017hyj,Visser:1989kh,Lobo:2005yv,Lemos:2003jb,Maldacena:2013xja,Hawking:1988ae,Sushkov:2005kj,Bronnikov:2002rn,Frolov:1990si,Delgaty:1994vp,Perry:1991qq,Cramer:1994qj,Oliva:2009ip,Maldacena:2004rf,Clement:1983fe,Clement:1995ya,Clement:1997yp,Guendelman:1991pc,Guendelman:2009er,Guendelman:2009pf,Guendelman:2008zp,Sakalli:2015taa,Sakalli:2015mka,Richarte:2017iit,Ovgun:2018xys,Ovgun:2018uin,Ovgun:2017jip,Ovgun:2016ijz,Ovgun:2015sqa,Jusufi:2018kmk,Jusufi:2017vta,Jusufi:2017mav,Halilsoy:2013iza,AzregAinou:1993ye,Gibbons:2017djb,Kim:2003zb}.

To detect the wormhole, a possible method is the application of optical gravitational lensing. The gravitational lensing by wormhole was studied widely in the literature of astrophysics as well as theoretical physics \cite{Kuhfittig:2013hva,Tsukamoto:2017hva,Tsukamoto:2016zdu,Tsukamoto:2016qro,Tsukamoto:2016oca,Tsukamoto:2016jzh,Sharif:2015qfa,Shaikh:2017zfl,Sajadi:2016hko,Lukmanova:2016czn,Kitamura:2016vad,Asada:2017vxl,Sakalli:2014bfa,Nandi:2016uzg,Kuhfittig:2015sta,Yoo:2013ukp,Tsukamoto:2012xs,Bhattacharya:2010zzb,Nandi:2006ds}. Using the Gauss-Bonnet theorem (GBT), Gibbons and Werner (GW) \cite{Gibbons:2008ru,Gibbons:2008rj} showed that it is possible to calculate deflection angle in weak limits, then Werner extended this method to Kerr black holes using the Nazim's osculating Riemannian method with
Randers-Finsler metric \cite{Werner:2012rc}. Afterwards,  using the finite distance from a lens object to a light source and a receiver, Ishihara, Suzuki, Ono and Asada \cite{Ishihara:2016sfv} calculated the deflection angle in a static, spherically symmetric and asymptotically flat spacetime,  and then recently extended by  Ono, Ishihara, and Asada (OIA) to calculate weak gravitational lensing in stationary axisymmetric spacetimes \cite{Ono:2017pie}. These methods use the GBT to calculate gravitational lensing that shows its global properties. The gravitational lensing effect, either in the weak gravitational field or in the strong gravitational field, it always requires the null geodesic equations.

Briefly in the GW method, it is considered a domain $\mathcal{D}_{R}$ bounded by the light ray  and a circular boundary curve $C_R$ centered on the lens which intersect the light ray at source $S$ and observer $O$, where both of them are at coordinate distance $R$ from the lens. For the asymptotic observer and source in the weak field approximation GW method demonstrates that when the GBT is used within the optical metric:
\begin{equation}
\iint\limits _{\mathcal{D}_{R}}K\,\mathrm{d}S+\oint\limits _{\partial\mathcal{D}_{R}}\kappa\,\mathrm{d}t+\sum_{i}\theta_{i}=2\pi\chi(\mathcal{D}_{R}),
\end{equation}
where  $dS$ is an areal element, $K$ is the optical Gaussian curvature,
the asymptotic deflection angle $\hat{\alpha}$ can be calculated by using the following equation: \cite{Gibbons:2008rj}
\begin{eqnarray} 
\hat{\alpha} & = & -\int\limits _{0}^{\pi}\int\limits _{r_{sl}}^{\infty}K \,\mathrm{d}S, \label{Alpha}
\end{eqnarray}
for the case of Euler characteristic $\chi(\mathcal{D}_{R}) =1$ and the summation of the jump angles $\sum_{i}\theta_{i}=\pi$.
Note that the integral is taken over the infinite region of the surface bounded by the light ray and excluding the lens. Furthermore, the photon orbit can be taken simply as the
straight line approximation $r_{sl}$ to calculate the leading term of $\hat{\alpha}$. Using the GW method or OIA method, a lot of studies of the weak gravitational lensing of light by black holes/wormholes have been done \cite{Jusufi:2017hed,Jusufi:2017lsl,Jusufi:2017mav,Jusufi:2017uhh,Jusufi:2017vew,Jusufi:2017vta,Jusufi:2018kmk,Ovgun:2018xys,Sakalli:2017ewb,Jusufi:2017xnr,Goulart:2017iko,Jusufi:2017gyu,Jusufi:2017drg,Ono:2017pie,Nakajima:2012pu,Crisnejo:2018uyn,Jusufi:2018jof}. 

Here, the main aim of the paper is to show that GBT and OIA method are valid for the calculating weak gravitational lensing by Damour-Solodukhin wormholes (DSW) which are the static Schwarzschild-like wormholes
solution recently found by Damour and Solodukhin in \cite{Damour:2007ap} and then rotating Kerr-like case is found by Bueno et al. in \cite{Bueno:2017hyj}. Moreover, we try to show the deflecting angle how much is shifted according to the parameter $\lambda$ from the Schwarzschild/Kerr black holes.  Moreover recently, it is shown that there is a relation between the strong gravitational lensing and the quasinormal modes (QNMs) in the context of black holes by Stefanov, Yazadjiev and Gyulchev \cite{Stefanov:2010xz}. For this purpose, we study the strong gravitational lensing by DSW using the method of Bozza \cite{bozza} who showed that the logarithmic divergence of the deflection angles at photon sphere exits and we show the relation of deflection angle with QNMs in strong regime.

The organization of the paper is as follows. In section 2 we briefly summarize the DSW, then we present the calculations of the weak gravitational lensing using the GBT and we calculate the strong gravitational lensing by the DSW. In section 3, we briefly give information about the rotating DSW and calculate the deflection angle of the rotating DSW using the OIA method. We conclude the paper in section 4.

\section{Damour-Solodukhin wormhole}

In this section, we consider the static Schwarzschild-like wormhole
solution, namely DSW \cite{Damour:2007ap} with metric:

\begin{equation}
ds^{2}=-(1-2M/r+\lambda^{2})dt^{2}+\frac{dr^{2}}{1-2M/r}+r^{2}d\Omega_{(2)}^{2}\,.\label{wormhole}
\end{equation}

 Note that this metric reduces to the Schwarzschild black hole at
$\lambda=0$. For non-zero values of the parameter $\lambda^{2}$,
the Einstein tensor of \ref{wormhole} has a zero $G_{tt}$, on the
other hand $G_{rr},G_{\theta\theta},G_{\phi\phi}\sim\lambda^{2}$
and need some matter to become toy model wormhole. Because of $t$ not correspond to the time of an asymptotic observer, we can
redefine the metric in \ref{wormhole} by using $t\rightarrow t/\sqrt{1+\lambda^{2}}$
and $M\rightarrow M(1+\lambda^{2})$:
\begin{equation}
ds^{2}=-f(r)dt^{2}+\frac{dr^{2}}{g(r)}+h(r) d\Omega_{(2)}^{2}\,,\label{wormhole2}
\end{equation}
where
\begin{equation}
f(r)=1-\frac{2M}{r}\,,\quad g(r)=1-\frac{2M(1+\lambda^{2})}{r}\,,
\end{equation}

and $h(r)=r^{2}$.

In the next subsection, we will study the weak gravitational lensing
using the Gauss-Bonnet theorem, and obtain the deflection angle in
weak limits.

\subsection{Weak gravitational lensing by Damour-Solodukhin wormhole using the
Gauss-Bonnet Theorem}

To calculate the deflection angle by DSW using the GBT \cite{Gibbons:2008rj}, we use the
equatorial plane $\theta=\pi/2,d\theta=0$ without losing generality,
due to the spherical symmetry, and the \ref{wormhole2} spacetime
reduces to orbital plane of light rays:
\begin{eqnarray}
ds^{2}=-f(r)dt^{2}+\frac{1}{g(r)}dr^{2}+r^{2}d\phi^{2}.\label{eq:metric2}
\end{eqnarray}
Using the two constants of motion in an affine parameter ($\lambda$):
\begin{eqnarray}
E=f(r)\frac{dt}{d\lambda},\quad L=r^{2}\frac{d\phi}{d\lambda},\label{eq:constant1}
\end{eqnarray}

where $E$ and $L$, are the energy and the angular momentum, respectively.
Then one can derive the another constant namely, impact parameter
$b=E/L$ as follows:
\begin{eqnarray}
b\equiv\frac{r^{2}\frac{d\phi}{d\lambda}}{f(r)\frac{dt}{d\lambda}},\label{eq:impact}
\end{eqnarray}
and the following relation is obtained $\frac{d\phi}{dt}=\frac{bf(r)}{r^{2}}.$ 

To define the optical metric $\bar{g}_{ij}$ which is also known as
the optical reference geometry ${\cal M}^{{\rm opt}}$,
we use the fact that each light ray satisfies the equation for null
geodesics $ds^{2}=0$, and the optical metric $\bar{g}_{ij}$
is written as follows:
\begin{eqnarray}
dt^{2}\equiv\bar{g}_{rr}dr^{2}+\bar{g}_{\phi\phi}d\phi^{2}=\frac{1}{f(r)g(r)}dr^{2}+\frac{r^{2}}{f(r)}d\phi^{2},\label{eq:optical1}
\end{eqnarray}

Afterwards, we use the slice of the constant time $t$ of the Eq.
(\ref{eq:metric2}), and we obtain a spatial part of spacetime in
two-dimensional curved subspace ${\cal M}^{{\rm sub}}$ as follows:
\begin{eqnarray}
d\ell^{2}\equiv g_{rr}dr^{2}+g_{\phi\phi}d\phi^{2}=\frac{dr^{2}}{g(r)}+r^{2}d\phi^{2}.\label{eq:metric3}
\end{eqnarray}

Next we use the conformal transformation with conformal factor $\omega^{2}(x)$
between Eq.s (\ref{eq:optical1}) and (\ref{eq:metric3}):
\begin{eqnarray}
\bar{g}_{\mu\nu}=\omega^{2}(x)g_{\mu\nu}.
\end{eqnarray}
It is noted that the conformal factor $\omega^{2}(x)=\frac{1}{f(r)}$,
and it does not change the condition of null geodesics.

It should be noticed that on the optical reference geometry ${\cal M}^{{\rm opt}}$,
$t$ plays the role of an arc length parameter because 
\begin{eqnarray}
\int_{t_{1}}^{t_{2}}dt=\int_{t_{1}}^{t_{2}}\sqrt{\bar{g}_{rr}(k^{r})^{2}+\bar{g}_{\phi\phi}(k^{\phi})^{2}}dt\\ =t_{2}-t_{1} \notag,\label{eq:optical3}
\end{eqnarray}

where the unit tangent vector $k^{i}$ of light ray paths on ${\cal M}^{{\rm opt}}$
as $k^{i}=\frac{dx^{i}}{dt}$ with the unit vector condition $1=\bar{g}_{ij}k^{i}k^{j}$.
Hence the GBT can be used on the optical reference geometry ${\cal M}^{{\rm opt}}$ as follows:
\begin{equation}
\iint\limits _{\mathcal{D}_{R}}K\,\mathrm{d}S+\oint\limits _{\partial\mathcal{D}_{R}}\kappa\,\mathrm{d}t+\sum_{i}\theta_{i}=2\pi\chi(\mathcal{D}_{R}),
\end{equation}

where $\mathcal{D}_{R}$ is a non-singular domain outside the light ray, within boundary $\partial\mathcal{D}_{R}=\gamma_{\tilde{g}}\cup C_{R}$, $\kappa$ stands for the geodesic curvature, $K$ is used for
the Gaussian curvature of optical metric, $\theta_{i}$ is the exterior
jump angles at the $i^{th}$ vertex, and  $\chi(\mathcal{D}_{R})=1$ is the Euler characteristic number.

The geodesic curvature $\kappa$ can be calculated with the following equation for the unit speed condition $\tilde{g}(\dot{\gamma},\dot{\gamma})=1$:
\begin{equation}
\kappa=\tilde{g}\,\left(\nabla_{\dot{\gamma}}\dot{\gamma},\ddot{\gamma}\right).
\end{equation}
When $R$ goes to infinity $R\rightarrow\infty$, the summation of the jump angles $\sum_{i}\theta_{i}$ are calculated as $\pi$ for the the source $\mathcal{S}$, and observer $\mathcal{O}$. Then the GBT is written as follows:
\begin{equation}
\iint\limits _{\mathcal{D}_{R}}K\,\mathrm{d}S+\oint\limits _{C_{R}}\kappa\,\mathrm{d}t\overset{{R\rightarrow\infty}}{=}\iint\limits _{\mathcal{D}_{\infty}}K\,\mathrm{d}S+\int\limits _{0}^{\pi+\hat{\alpha}}\mathrm{d}\phi=\pi,
\end{equation}

where the $K$ is the Gaussian curvature (gives information about how surface is curved) and the $K$ is defined as follows:
\begin{equation}
\begin{split}
&K=\\
&\frac{-1}{\sqrt{\bar{g}_{rr}\bar{g}_{\phi\phi}}}\left[\frac{\partial}{\partial r}\left(\frac{1}{\sqrt{\bar{g}_{rr}}}\frac{\partial\sqrt{\bar{g}_{\phi\phi}}}{\partial r}\right)+\frac{\partial}{\partial\phi}\left(\frac{1}{\sqrt{\bar{g}_{\phi\phi}}}\frac{\partial\sqrt{\bar{g}_{rr}}}{\partial\phi}\right)\right].\label{eq:GB2}
\end{split}
\end{equation}

Then the Gaussian curvature for the optical metric of DSW in \ref{eq:optical1}  is calculated:

\begin{equation}
K={\frac{\left(6\,{\lambda}^{2}+6\right){M}^{3}+\left(-7\,{\lambda}^{2}-7\right)r{M}^{2}+{r}^{2}\left({\lambda}^{2}+2\right)M}{\left(-r+2\,M\right){r}^{4}}}.\label{gcurv}
\end{equation}

The Gaussian curvature in \ref{gcurv} reduces to in this form up to leading orders:
\begin{equation}
\begin{split}
K=6\,{\frac{{M}^{3}}{\left(-r+2\,M\right){r}^{4}}}-7\,{\frac{{M}^{2}}{\left(-r+2\,M\right){r}^{3}}}+2\,{\frac{M}{\left(-r+2\,M\right){r}^{2}}}\\ +{\frac{M\left(6\,{M}^{2}-7\,Mr+{r}^{2}\right){\lambda}^{2}}{\left(-r+2\,M\right){r}^{4}}}+\mathcal{O}(\lambda^{3}).\label{Curvature}
\end{split}
\end{equation}

Afterwards, we calculate the geodesic curvature $\kappa$ which shows how far the curve
$C_{R}$ deviates from the geodesic, using the following equation:
\begin{eqnarray}
\kappa=\frac{1}{2\sqrt{\bar{g}_{rr}\bar{g}_{\phi\phi}}}\left(\frac{\partial\bar{g}_{\phi\phi}}{\partial r}\frac{d\phi}{dt}-\frac{\partial\bar{g}_{rr}}{\partial\phi}\frac{dr}{dt}\right).\label{eq:GB3}
\end{eqnarray}
Note that if the trajectory of light ray $\gamma$ is geodesic, the geodesic curvature is zero $\kappa(\gamma)=0$ so that we can choose $C_{R}:=r(\phi)=R=\text{const}$. At $R$ goes to $\infty$, the geodesic curvature $\kappa$ reduces to \begin{eqnarray}
\lim_{R\rightarrow\infty}\kappa(C_{R})& \rightarrow & \frac{1}{R}.
\end{eqnarray} 
Additionally,  at $R$ goes to $\infty$, optical metric also goes to:
\begin{eqnarray}
\lim_{R\rightarrow\infty}\mathrm{d}t  & \to & R\,\mathrm{d}\varphi.
\end{eqnarray}

Using the straight line approximation of the light ray as $r=b/\sin\varphi$, the deflection angle by the DSW can be calculated using the GBT as follows:
\begin{eqnarray}
\hat{\alpha} & = & -\int\limits _{0}^{\pi}\int\limits _{\frac{b}{\sin\varphi}}^{\infty}K \,\mathrm{d}S, \label{int1}
\end{eqnarray}
where 
$dS=\sqrt{\det|\bar{g}|}drd\phi$
is an areal element and
the $\kappa(C_{R})\mathrm{d}t=\mathrm{d}\,\varphi$ is used.

Using the Gaussian curvature $K$ Eq. \eqref{Curvature} into the above integral, the deflection angle by DSW within the leading order terms  (weak lensing) for the asymptotic source and receiver is calculated as follows:
\begin{equation}
\hat{\alpha}\simeq\,{\frac{4M}{b}} +{\frac{2M{\lambda}^{2}}{b}} 
\,\label{GB1}.
\end{equation}

The deflection angle by DSW is increased with the ratio of the parameter $\lambda$ as seen in the Eq. (\ref{GB1}) with compared to Schwarzschild black hole \cite{Gibbons:2008rj}.

\subsection{Strong deflection limit of Damour-Solodukhin wormhole and its relation with QNMs}

Using the Bozza's procedure \cite{bozza}, we study the strong gravitational lensing (SGL) of the DSW in the case of photons passing very close to the photon sphere, with radius $r_{m}$. We use the assumption ($\theta=\pi/2$), with the light ray's trajectory
\begin{equation}
\dot{r}^{2}=g(r)\left( \frac{E^{2}}{f(r)}-\frac{L^{2}}{h(r)}\right)=0,
\end{equation}
where "dot" is for the derivative respect to an affine parameter. Note that the conserved energy is $E\equiv f(r)\dot{t}>0$ whereas the angular momentum defined as $L\equiv r^{2}(r)\dot{\phi}$ in Eq. (\ref{eq:constant1}). From the null circular orbit,  we can find by the largest positive solution of the equation, 
\begin{equation}
\frac{h(r)'}{h(r)}=\frac{f(r)'}{f(r)}.\label{d11}
\end{equation}

Using the Eq.~(\ref{d11}), it is found that $r_{m}=3M$ and $r_{m}>r_{throat}$ is satisfied. Moreover, light ray is deflected from the closest approach distance of the photon $r_{c}$ smaller than $r$ so that this condition should be considered where the light ray is supposed to come from infinity and deflect by DSW. Using the conservation of the angular momentum, the closest approach distance $r_{c}$ which is related to the impact parameter is calculated as 
\begin{equation}
u=\sqrt{\frac{h(r_c)}{f(r_c)}}=\sqrt {{{{\it r_c}}^{2} \left( 1-2\,{\frac {M}{{\it r_c}}} \right) ^{-1}}}.\end{equation}

After we use the definition of the critical impact parameter $u_{cr}$ at strong deflection limit $r_{c}\rightarrow r_{m}$ or $u\rightarrow u_{m}$  as 
\begin{equation}\label{eq:critical_impact_parameter1}
u_{m}(r_{m})
\equiv \lim_{r_{c}\rightarrow r_{m}} \sqrt{\frac{h(r_c)}{f(r_c)}},
\end{equation}
we obtain :
\begin{equation}
u_{m}=3\,\sqrt {3}{M}.
\end{equation}

Then we calculate the exact deflection angle $\alpha$ for the DSW as follows:

\begin{eqnarray}\label{eq:da}
\alpha(r_{c}) =I(r_{c})-\pi,
\end{eqnarray}
where $I(r_{c})$ is calculated by
\begin{equation}
I =2 \int^{\infty}_{r_{c}}\frac{ dr}{\sqrt{g(r) h(r) }\sqrt{\frac{h(r) f(r_c)}{h(r_c) f(r)}-1}}. \label{IR}
\end{equation}

Note that when $u$ decreases, the bending angle $\alpha$ increases that the light rays encircle the DSW completely till $2 \pi$. At $r_{c}=r_{m}$ due to $u = u_{m}$ the photon will be trapped inside an orbit. As it diverges  in the SGL $u\rightarrow u_{m}$ or $r_c\rightarrow r_m$, we rewrite the deflection angle in this form which is used for ultra static spacetimes:
\begin{equation}\label{eq:defstr}
\alpha(u)=-a\log \left( \frac{r_{c}}{r_{m}}-1 \right) +b+O((r_{c}-r_{m})\log (r_{c}-r_{m})),
\end{equation}
where $a$ and $b$ are SGL constants.
The impact parameter $u$ is also related to angular separation of the image from the lens: $\theta=\frac{u}{D_{OL}}$, where $D$ is the distance between the lens
$D_{OL}$ and the observer. Then the strong field limit of the deflection angle can be calculated as follows:

\begin{equation}\label{eq:defstr2}
\alpha(u)=-\bar{a}\log \left( \frac{\theta D_{OL}}{u_{m}}-1 \right) +\bar{b} +O[(u-u_{m})\log (u-u_{m})],
\end{equation}
with

\begin{eqnarray}
\overline{a}&=& \sqrt{{2 f(r_m) \over g(r_m) \left[ h''(r_m) f(r_m)-h(r_m) f(r_m)''\right]}}\label{a} \label{up} \notag  \\
 &=&-{\frac{1}{\sqrt {1-2\,{\lambda}^{2}}}},
\end{eqnarray}

\begin{equation}\label{eq:bbar1}
\overline{b}=\overline{a}\log \left[r^{2}_{m}\left(\frac{h''(r_m)}{h(r_m)}-\frac{ f''(r_m)}{ f(r_m)}\right)\right] +I(r_{m})-\pi 
\end{equation}

\begin{eqnarray}
={\frac{ ln(6) \sqrt {-2\,{\lambda}^{2}+1}}{ \,{ ln(10) 2\lambda}^{2}-1}}+{
\frac{ 2\left( {\lambda}^{2}-1/2 \right)  \left( {\it I (r_m)}-\pi \right) 
}{2\,{\lambda}^{2}-1}}, \notag
\end{eqnarray}

where expression for $I(r_m)$ can be found in \cite{bozza} (only solve numerically). Note that prime denotes the derivative with respect to the radial coordinate $r$. Moreover, all the information about the SGL is encoded into $\bar{a}$ and $\bar{b}$. Then we find the relation with the QNM of DSW \cite{Stefanov:2010xz}:

\begin{equation}\label{QNMLyapunov}
\omega_{\rm QNM}=\Omega_p\,l-i(n+1/2)\,|\Lambda|\,,
\end{equation}
where \begin{equation}\Lambda =c\,\sqrt{\frac{g(r_m)\left[f(r_m) h''(r_m)-f''(r_m) h(r_m))\right]}{2 h(r_m)}} \end{equation}
\begin{equation}
\Lambda=\,{\frac {c\sqrt {3-6\,{\lambda}^{2}}}{9M}}, \notag
 \end{equation}
with speed of light $c$.
The parameter $\Lambda$ which appears in the imaginary part is
the Lyapunov exponent which determines the instability timescale of the orbit. Then the simple relation can be written as follows:
\be
\Lambda={c\over u_m \overline{a}}.\label{relat_lambda}
\ee

The other important relation is the coordinate angular velocity with the impact parameter of the lens
\be
\Omega_m= c\,\sqrt{f(r_m)g(r_m)}={c\over r_m}\label{relat_omega}.
\ee
After using the equations (\ref{relat_lambda}) and (\ref{relat_omega}), it is easily observed that

\be
\overline{a} = \frac{\Omega_p}{\Lambda}\label{lambda_omega}.
\ee

\section{Rotating Damour-Solodukhin Wormhole}

In this section, we have briefly describe rotating DSW spacetime. The Kerr-like wormhole metric is constructed using the method of Damour and Solodukhin \cite{Bueno:2017hyj}:
\begin{align}
 & ds^{2}=-\left(1-\frac{2Mr}{\Sigma}\right)dt^{2}-\frac{4Mar\sin^{2}\theta}{\Sigma}dtd\phi+\frac{\Sigma}{\hat{\Delta}}dr^{2}\nonumber \\
 & +\Sigma d\theta^{2}+\left(r^{2}+a^{2}+\frac{2Ma^{2}r\sin^{2}\theta}{\Sigma}\right)\sin^{2}\theta d\phi^{2}\,,\label{keeerr}
\end{align}
which \begin{equation}
\Sigma\equiv r^{2}+a^{2}\cos^{2}\theta,\quad\hat{\Delta}\equiv r^{2}-2M(1+\lambda^{2})r+a^{2}.
\end{equation}

Note that the $M$ is the mass and  the $aM$ stands for the angular momentum. For the case of $\lambda^{2}=0$, we can recover the Kerr metric, on the other hand, for non-vanishing $\lambda^{2}$,
the structure of the spacetime is totally changed. We have calculated the positive root
of $\hat{\Delta}$: $r_{+}=(1+\lambda^{2})M+\sqrt{M^{2}(1+\lambda^{2})^{2}-a^{2}}$ that gives special surface, but not  the surface of the event horizon, and the throat of the wormhole is located at $r=r_{+}$. For the non vanishing values of the $\lambda\neq0$, the Kerr-like wormhole is constructed and its QNMs are recently studied in \cite{Bueno:2017hyj}.

Now, we will study the weak gravitational lensing
by rotating DSW with OIA method, and obtain the deflection angle in
weak limits.

\subsection{Deflection angle of rotating Damour-Solodukhin Wormhole using OIA method}
Here, we use the OIA method to investigate the weak gravitational lensing by rotating DSW, known as Kerr-like wormhole \cite{Ono:2017pie}.
First, using the null condition ($ds^2 = 0$) in the rotating DSW spacetime given in Eq. \ref{keeerr}, we obtain the generalized
optical metric $\gamma_{ij}$ ($i, j = 1, 2, 3$) in this form
{\cite{Ono:2017pie,Ono:2018ybw}}:

\begin{align}
  dt = & \sqrt{\gamma_{ij} dx^i dx^j} + \beta_i dx^i, \label{opt}
\end{align}

with the components

\begin{align}
\gamma_{ij}dx^idx^j=&
\frac{\Sigma^2}{\quad\hat{\Delta}(\Sigma-2Mr)}dr^2 
+ \frac{\Sigma^2}{(\Sigma-2Mr)} d\theta^2
\notag\\
&+ \left(r^2+a^2+\frac{2a^2Mr\sin^2\theta}{(\Sigma-2Mr)}\right) 
\frac{\Sigma\sin^2\theta}{(\Sigma-2Mr)} d\phi^2 ,
\label{gamma}
\\
\beta_idx^i=&- \frac{2aMr\sin^2\theta}{(\Sigma-2Mr)}d\phi .
\label{beta-Kerr}
\end{align}

Then we calculate the Gaussian optical curvature $K$ \ in the weak
field approximation   as follows:

\begin{equation}
\begin{split}
  K \approx -{\frac { \left( {\lambda}^{2}+2 \right) M}{{r}^{3}}}.
\end{split}
\end{equation}
Then, one should use the OIA method with the straight line approximation of the light ray defined as $r=b \left( \sin \left( \phi \right) +{\frac {M \left( 1+ \left( \cos
 \left( \phi \right)  \right) ^{2} \right) }{b}}-2\,{\frac {aM}{{b}^{2
}}} \right) ^{-1}$ and the geodesic
curvature to find the contribution of rotating term $aM$ in deflection angle (for details see the method by Ono, Ishihara and Asada in Ref. \cite{Ono:2017pie}).

\begin{figure}[!ht]

\includegraphics[width=0.4\textwidth]{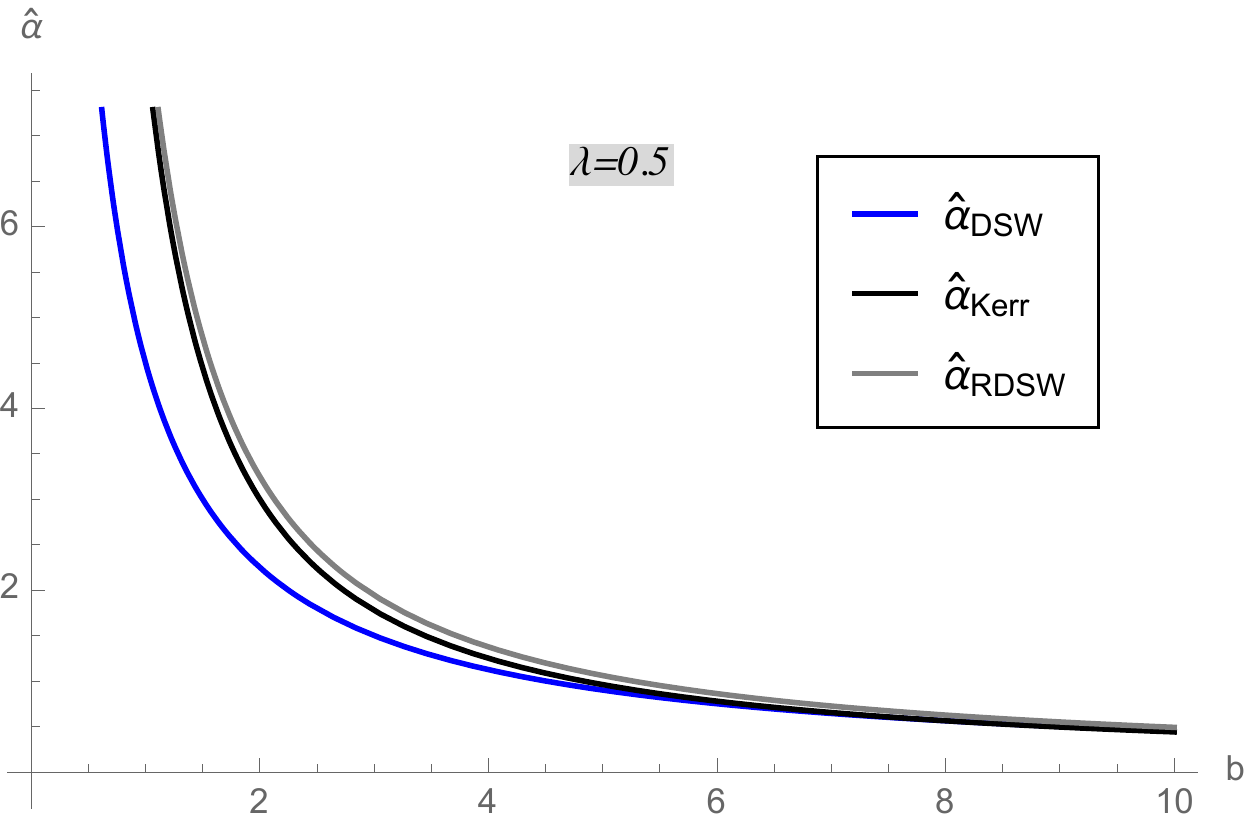}
\includegraphics[width=0.4\textwidth]{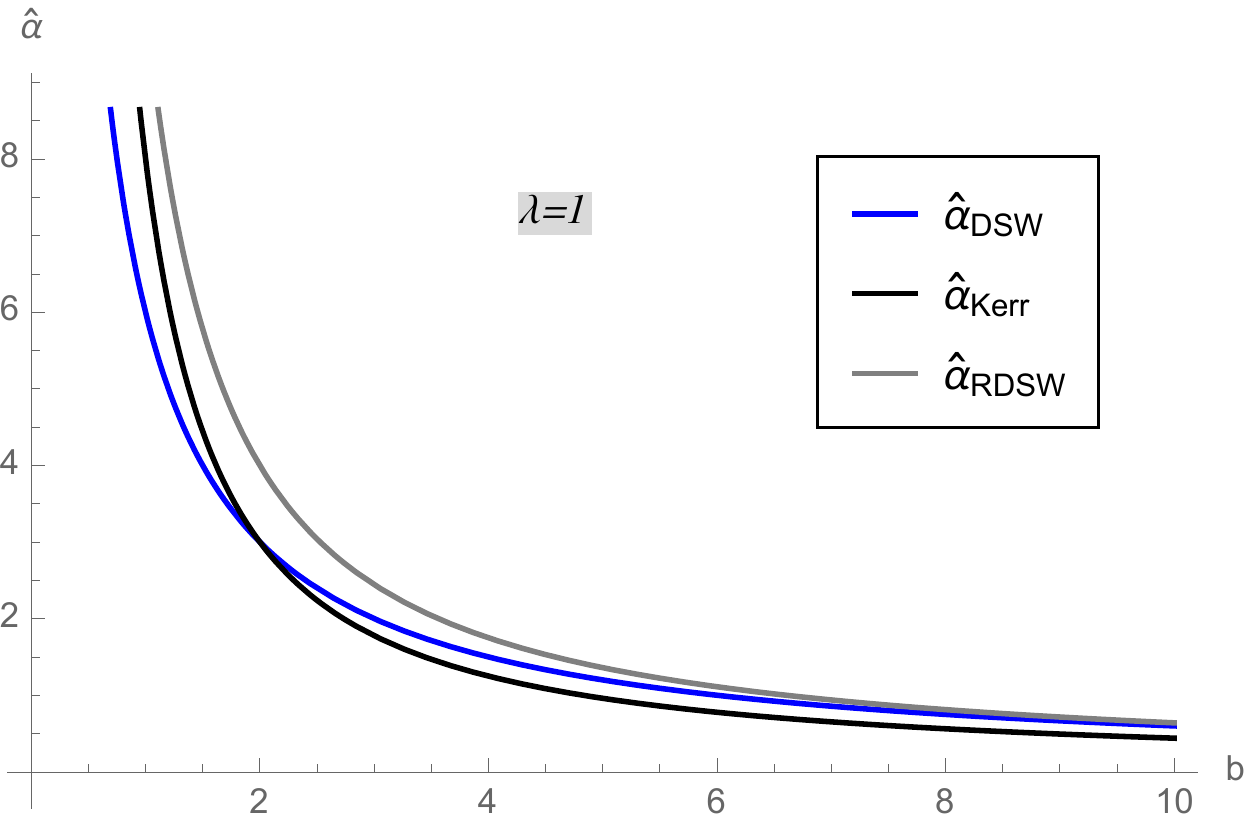}
\includegraphics[width=0.4\textwidth]{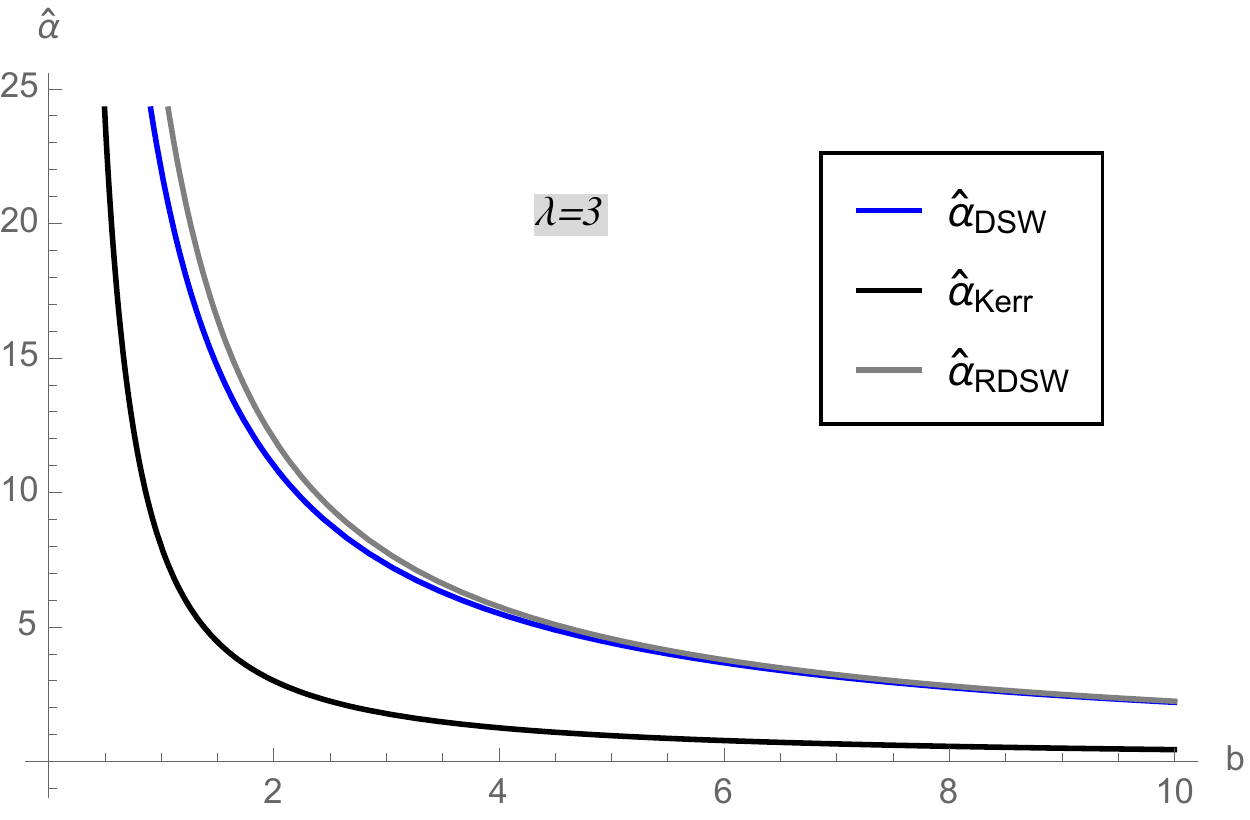} 
\includegraphics[width=0.4\textwidth]{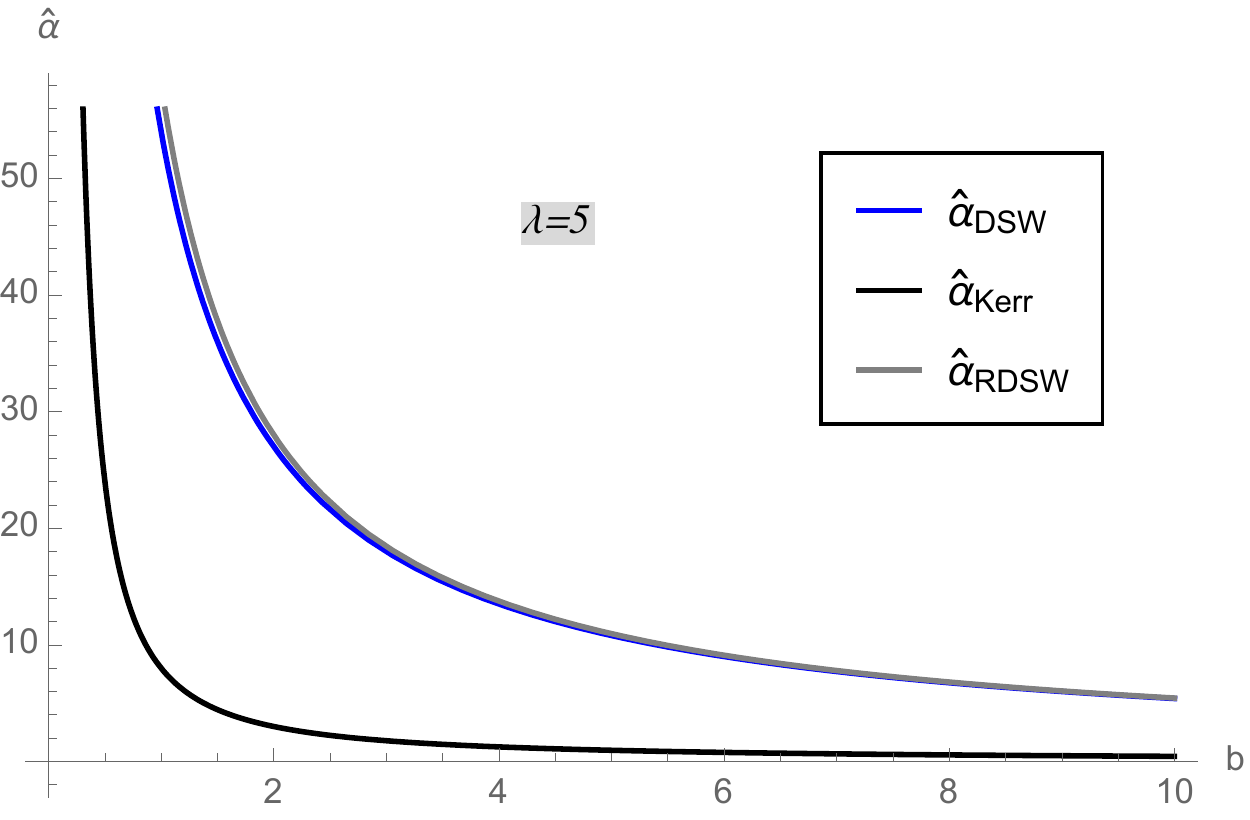}
 \caption{ The deflection angles by DSW, Kerr black hole, and rotating DSW (RDSW) are plotted as a function of the $b$ for $M=a=1$ and four different values of $\lambda$: $\lambda=0.5$, $\lambda=1$, $\lambda=3$ and $\lambda=5$.} \label{Fig1}
\end{figure}

Afterwards we obtain the geodesic curvature $\kappa$ on the equatorial plane for the slowly rotating DSW in weak field limit \cite{Ono:2017pie,Ono:2018ybw}:

\begin{equation}
\kappa=-\sqrt{\frac{1}{\gamma \gamma ^{\theta \theta }}}\beta _{\phi
,r} \label{kappa}
\end{equation}

\begin{align}
\kappa=&-\frac{2aM}{r^3}+\,{\frac {{2M}^{2}a{\lambda}^{2}}{{r}^{4}}}-
\,{\frac {2a{M}^{2}}{{r}^{4}}}.
\label{kappag-second}
\end{align}
Then the net contribution of the geodesic part at infinite distance limit becomes: 

\begin{equation}
\int \kappa d\ell =\pm \,{\frac {{4M}\,a}{{b}^{2}}},  \label{intkappa}
\end{equation}
where the positive sign stands for a retrograde light rays and negative sign is for the prograde light rays. Note that there is not any $\lambda$ term in geodesics part, because we consider only leading order term of $M$ in weak field limit.
Hence, we calculate the total
deflection angle of the rotating DSW at the leading order of the weak field approximation:
\begin{equation}
  \hat{\alpha} =\,{\frac {2M \left( {\lambda}^{2}+2 \right) }{b}} \pm  \,{\frac {{4M}\,a}{{b}^{2}}}.
 \label{deflectionangle2}
\end{equation}
We note that only the first order terms in $aM$ should be correct. One can modify the integration domain $S_\infty$, to obtain the correct second order terms as well similarly to paper of \cite{Ono:2017pie}. In \ref{deflectionangle2} first term stands for the static wormhole while the second terms lead the contribution of the rotation. 
For the case of $a$ goes to zero, the total deflection angle of rotating DSW reduces to the nonrotating DSW case which we found in \ref{GB1}. Furthermore, for the case of $\lambda=0$, the deflecting angle of the  Kerr black hole is recovered \cite{Ono:2017pie}.  We plot the deflection angles by DSW, Kerr black hole, and rotating DSW (RDSW) as a function of the closest approach distance $b$ for $M=a=1$ and four different values of $\lambda$: $\lambda=0.5$, $\lambda=1$, $\lambda=3$ and $\lambda=5$ to illustrate the effect of the wormhole parameter $\lambda$ in Fig. \ref{Fig1}.

\section{Conclusions}

In summary, the observation of wormholes by studying the gravitational lensing is one of the most effective way to testify them in the universe.

For this purpose, first, we have explicitly calculated the deflection angle of the light by DSW in the weak field approximation using the method developed by GW.  

Second, we have studied the strong gravitational lensing using the method of Bozza, and show the relation of the strong deflection angle with its QNMs. 

Last, we have briefly described the DSW in a rotating Kerr-like black hole spacetime and studied the weak gravitational lensing using OIA method where the receiver and source are at the null infinity. There are two methods to calculate the deflection angle of rotating black holes in GBT. First method is the Nazim's osculating Riemannian method with
Randers-Finsler metric which is found by Werner  \cite{Werner:2012rc}. In this paper, we have used the second method recently found by Ono, Ishihara and
Asada (OIA) in Ref. \cite{Ono:2017pie}. A huge amount of
calculations is needed in the method of Werner for arriving at Eq. (\ref{deflectionangle2}), on the other hand, the method proposed by OIA in Ref. \cite{Ono:2017pie} enables us to do more easily the calculations.

The significant of this result is that the deflection of a light ray is calculated by outside of the lensing region which means that the effect of the gravitational lensing is a global effect such that there are more than one light ray converging between the source and observer. Hence, we are able to find accurately deflecting angle in weak-field limits. We finally conclude that the deflection angle by the DSW is increased with the DSW parameter of $\lambda$ in Fig. \ref{Fig1}.  

With regards to future work, it would be interesting to see whether this approach could also be extended to the other compact objects and also see whether there is an effect of dark matter on the deflection angle. 
A careful studies on the gravitational lensing and also gravitational waves with QNMs may shed some light on possible signature of the existence of the wormholes.

\acknowledgments The authors gratefully thank to the Referees for the constructive comments and recommendations which definitely help to improve the readability and quality of the paper. This work was supported by the Chilean FONDECYT
Grant No. 3170035 (A. \"{O}.).  A. \"{O}. is grateful to Prof. Douglas Singleton for hosting him at the California State University, Fresno and also thanks to Prof. Leonard Susskind and Stanford Institute for Theoretical Physics for hospitality. A. \"{O}. would like to thank Institute for
Advanced Study, Princeton for hospitality.

\end{document}